## Preprint notes

**Title of the article:**

Time for AI (Ethics) Maturity Model Is Now

**Authors:**

Ville Vakkuri, Marianna Jantunen, Erika Halme, Kai-Kristian Kemell, Anh Nguyen-Duc, Tommi Mikkonen, Pekka Abrahamsson

**Notes:**

-This is the author's version of the work

-This is a pre-print of an article accepted to the Workshop on Artificial Intelligence Safety (SafeAI 2021)
https://safeai.webs.upv.es/

# Time for AI (Ethics) Maturity Model Is Now

Ville Vakkuri[1], Marianna Jantunen[1], Erika Halme[1],
Kai-Kristian Kemell[1], Anh Nguyen-Duc[2], Tommi Mikkonen[3], Pekka Abrahamsson[1]

[1]University of Jyväskylä, Faculty of Information Technology
ville.vakkuri — marianna.s.p.jantunen — erika.a.halme— kai-kristian.o.kemell — pekka.abrahamsson@jyu.fi
[2]University of South Eastern Norway, School of Business
angu@usn.no
[3]University of Helsinki, Department of Computer Science
tommi.mikkonen@helsinki.fi

**Abstract**

There appears to be a common agreement that ethical concerns are of high importance when it comes to systems equipped with some sort of Artificial Intelligence (AI). Demands for ethical AI are declared from all directions. As a response, in recent years, public bodies, governments, and universities have rushed in to provide a set of principles to be considered when AI based systems are designed and used. We have learned, however, that high-level principles do not turn easily into actionable advice for practitioners. Hence, also companies are publishing their own ethical guidelines to guide their AI development. This paper argues that AI software is still software and needs to be approached from the software development perspective. The software engineering paradigm has introduced maturity model thinking, which provides a roadmap for companies to improve their performance from the selected viewpoints known as the key capabilities. We want to voice out a call for action for the development of a maturity model for AI software. We wish to discuss whether the focus should be on AI ethics or, more broadly, the quality of an AI system, called a maturity model for the development of AI systems.

## Introduction

The ethics of Artificial Intelligence (AI) have been an emerging topic in the field of AI development (Jobin, Ienca, and Vayena 2019), and the ethical consequences of AI systems have been researched in significant amounts in the recent years (Ryan and Stahl 2020). Now that AI has become prevalent in many decision-making processes that have the chance to directly or indirectly impact or alter lives, in fields such as healthcare (Panesar 2019) and transportation (Sadek 2007), concerns regarding the currently existing and hypothetical ethical impacts of AI systems have been voiced by many. With AI systems becoming pervasive, there emerges an increasing need for guidance in creating AI systems that align with our perception of ethical behavior.

As Jobin, Ienca, and Vayena (2019) suggest, there is an apparent agreement that AI should be ethical. Still, the details of what "ethical AI" constitutes and "which ethical requirements, technical standards and best practices are needed for its realization", is up for debate. The ethics of AI systems appear open for initiatives, or as Greene, Hoffmann, and Stark (2019) put it, 'up for grabs'. These initiatives offer goals and definitions for what is expected of ethical AI systems. As stated in Ethically Aligned Design: A Vision for Prioritizing Human Well-being with Autonomous and Intelligent Systems, First Edition, regardless of the ethical framework we follow, the systems should be expected to honor holistic definitions of societal prosperity, not pursuing one-dimensional goals such as increased productivity or gross domestic product. Awad et al. (2018) proposed that we were entering an era where intelligent systems can be tasked to "not only to promote well-being and minimize harm, but also to distribute the well-being they create, and the harm they cannot eliminate". Societal and policy guidelines should be established to ensure that they remain human-centric, serving humanity's values and ethical principles (Ethically Aligned Design: A Vision for Prioritizing Human Well-being with Autonomous and Intelligent Systems, First Edition).

Coming up with policies and enforcing them within an organization might seem challenging and unrewarding. Demands for ethical AI are declared from all directions, but the rewards and consequences of making or not making ethical initiatives and commitments seem unclear. When companies and research institutions make "ethically motivated 'self-commitments'" in the AI industry, efforts to formulate a binding legal framework are discouraged, and any demands of AI ethics laws remain relatively vague and superficial (Hagendorff 2020). As Greene, Hoffmann, and Stark (2019) suggest, many high-profile companies, organizations, and communities have signaled their commitment to ethics, but the resulting articulated value statements "prompt more questions than answers". A problem may also emerge in the situation where – as presented by Hagendorff (2020) – AI ethics, like ethics in general, "lacks mechanisms to reinforce its own normative claims". It might be that the consequences of not enforcing and applying ethical principles in AI development are not severe enough to motivate companies to follow through.

Despite these challenges, many organizations have reacted to ethical concerns on AI, for example, by forming ad-hoc expert committees to draft policy documents (Jobin, Ienca, and Vayena 2019) and producing statements that describe ethical principles, values and other abstract requirements for AI development and deployment (Mittelstadt 2019). At least 84 public-private AI ethics principles and values initiatives were identified by Mittelstadt (2019), and the topic evolves dynamically through new initiatives and their iterations. Such initiatives can "help focus public debate on a common set of issues and principles, and raise awareness among the public, developers and institutions of the ethical challenges that accompany AI" (Mittelstadt 2019; Whittaker et al. 2018).

So far, these principles and values used to form various guidelines for implementing AI ethics have been the primary tools intended to help companies develop ethical AI systems (as we discuss in detail in the next section). However, as already noted in existing literature (Mittelstadt 2019), these guidelines alone cannot guarantee ethical AI systems, and seem to suffer from a lack of industry adoption (Vakkuri et al. 2020). What, then, should be done instead? In this paper, we look at the issue from the point of view of Software Engineering (SE).

One approach to tackling this issue, from the point of view of SE, would be to focus on methods, practices, and tools for AI ethics, in order to make these principles and values more tangible to the developers working on these AI systems. Some already exist, as discussed by Morley et al. (2019), although they are mostly technical ones focused specifically on, e.g., managing some aspects of machine learning. Another approach, on which we focus here, is the development of a maturity model. Maturity models, which we discuss further in the third section, are used in SE to evaluate the maturity level of organizational processes related to software development. Could a maturity model for AI ethics help organizations develop ethical AI?

## AI Ethics Guidelines

To respond to the concerns and discussions around the ethical and societal impacts of intelligent technology, guidelines for ethical AI development have been published in the recent years by a variety of organizations ranging from corporations to governmental and research institutions. Still, there appears to be no acknowledged single standard in the field, but the guidelines often appear to be either one "keyword" principles such as *accountability* or *transparency* (Ethically Aligned Design: A Vision for Prioritizing Human Well-being with Autonomous and Intelligent Systems, First Edition) or descriptive sentences that present the organization's approach, such as "We want to develop safe, robust, and explainable AI products" (Bolle 2020). The guidelines may serve different purposes for each organization – a corporation's motivation to publishing a set of ethical guidelines to follow can be expected to be different from that of a research institution.

Tending to the need of standards, several organizations have stepped in to publish their own guidelines. As phrased by Fjeld et al. (2020), "seemingly every organization with a connection to technology policy has authored or endorsed a set of principles for AI". As an example of the aforementioned policy forming committees (Jobin, Ienca, and Vayena 2019), some major publications from influential institutions, such as The IEEE Ethically Aligned Design: A Vision for Prioritizing Human Well-being with Autonomous and Intelligent Systems, First Edition; and Ethics Guidelines for Trustworthy AI by the High-Level Expert Group appointed by the European Commission, have introduced practical design approaches and suggested standards and principles for ethical AI development and implementation. Research institutions are only the tip of the iceberg, however; a variety of other institutions, such as governments and corporations, have stepped in to publish their own AI ethics guidelines, as discovered by, for example, Jobin, Ienca, and Vayena (2019). Even the Vatican has published their initiative, teaming up with IBM and Microsoft to draft a call for AI ethics (Stotler 2020).

While not legally binding, the effort invested in such guidelines by multiple stakeholders in the field are noteworthy and influential (Jobin, Ienca, and Vayena 2019), contributing to the discussion of AI ethics. Guidelines can be seen as "part of a broader debate over how, where, and why these technologies are integrated into political, economic, and social structures" (Greene, Hoffmann, and Stark 2019) (p. 2122). We can witness how guidelines have contributed positively to the development of AI ethics discussion by observing the number of organizations that published their sets of guidelines. Based on the number of organizations that use the common vocabulary of "keyworded" guidelines, discussing transparency, fairness, and other such principles, it seems as though guidelines may have developed into a type of "common language" for AI ethics discussion; a familiar format that is easy to adopt and quick to communicate.

Researchers have conducted reviews on AI ethics guidelines, considering their implications (e.g. Ryan and Stahl (2020)) and looking for unanimity among them (e.g. Jobin, Ienca, and Vayena (2019); Hagendorff (2020)). In the light of these reviews, certain prevalent guidelines have emerged. For example, Jobin, Ienca, and Vayena (2019) identified a "global convergence emerging around five ethical principles", namely transparency, justice and fairness, non-maleficence, responsibility and privacy.

However, guidelines alone do not cater to the whole spectrum of AI ethics challenges. Firstly, although some similarities emerge between sources and studies, there is no guarantee of unanimity of their application; even if each organization were to adhere to the exact same set of guidelines, their practical application is not guaranteed to be synchronized. There may be questions related to, for example, interpretation, emphasis and level of commitment, that organizations need to make for themselves.

In particular, when considering organizations employing guidelines in their AI product development, the guidelines often provide us with the answer to the question "what" is done, but not "how". This concept seems supported by Morley et al. (2020) when they discuss the same effect on the mainstream ethical debate on AI. Another problem following from reliance in guidelines is, that their impact on human decision-making is not guaranteed, and they may remain ineffective (Hagendorff 2020).

As reported by Vakkuri et al. (2019), there appears to be a gap between research and practice in the field of AI ethics when it comes to the procedures of companies, as the academic discussions have not carried over to industry; developers consider ethics important in principle, but perceive them to be distant from the issues they face in their work. In a survey of industry practices, including 211 companies, 106 which develop AI products, it was found that companies have mixed levels of maturity in implementing AI ethics (Vakkuri et al. 2020). In terms of guidelines, the survey discovered that the various AI ethics guidelines had not, in fact, had a notable effect on industry practices, confirming the suspicions of Mittelstadt (2019).

The high variety in both industry practices and AI ethics guidelines may make it difficult to assess AI systems development, especially aspects such as trustworthiness or other ethics-related topics. To answer a need of standardized evaluation practices, we propose a look into maturity models, and their utility in evaluating software development practices. Maturity models or maturity practices for AI with different emphases have already been introduced, such as the AI-RFX Procurement Framework by The Institute for Ethical AI and Machine Learning (The Institute for Ethical AI and Machine Learning 2020) and The AI Maturity Framework (Ramakrishnan et al. 2020). Next, we discuss maturity models in general, before discussing them further in the specific context of AI and AI ethics in the fourth section.

## What are Maturity Models?

Maturity models are intended to help companies appraise their process maturity and develop it. They serve as points of reference for different stages of maturity in an area. In the context of SE, they are intended to help organizations move from ad hoc processes to mature and disciplined software processes (Herbsleb et al. 1997). Since the Software Engineering Institute launched the Capability Maturity Model (CMM) almost twenty years ago (Paulk et al. 1993), hundreds of maturity models have been proposed by researchers and practitioners across multiple domains, providing framework to assess current effectiveness of an organization and supports figuring out what capabilities they need to acquire next in order to improve their performance.

Though maturity models are numerous in SE, the Scaled Agile Framework (SAFe) and Capability Maturity Model Integration (CMMI) are some typical high-profile examples of maturity models in the field of SE. SAFe is a mixture of different software development practices and focuses mainly on scaling agile development in larger organizations. CMMI, on the other hand, focuses on improvements related to software development processes. In general, Software Process Improvement tools are rooted in Shewhart-Deming's plan-do-check-act (PDCA) paradigm, where CMMI, for example, represents a prescriptive framework in which the improvements are based on best practices (Pernstål et al. 2019).

Maturity models have been studied in academic research as well. Studies have focused on both their benefits and the potential drawbacks. For example, a past version of the CMMI has been criticized for creating processes too heavy for the organizations to handle (Sony 2019; Meyer 2013), and in general being resource-intensive to adopt for smaller organizations (O'Connor and Coleman 2009). SAFe, on the other hand, has been criticized for adding bureaucracy to Agile (Ebert and Paasivaara 2017), leaning towards the waterfall approach.

Nonetheless, these models are widely used in the industry, either independently, or in conjunction with other frameworks, tools, or methods. SAFe, for example, has been adopted by 70 of the Forbes 100 companies. CMMI has even been adopted in fields other than software development. Academic studies aside, companies seem to have taken a liking to maturity models in the context of software.

Indeed, this apparent popularity of these models out on the field has, in part, motivated us to write this early proposal maturity models in the context of AI ethics as well. In an area where we struggle with a gap between research and practice, we argue that looking at frameworks, models, and other tools that are actively used out on the field is a good starting point for further steps. Thus far, guidelines have been used to make AI ethics principles more tangible, but further steps are still needed, and a maturity model could be one such step.

## What about an AI Ethics Maturity model or an AI Maturity Model?

Despite the criticism towards maturity models discussed above, maturity models are widely used in the industry. Conversely, the AI ethics guidelines that have been somewhat well-received in the academia seem to not have seen much interest out on the field. We thus propose that an AI development maturity model might take us closer to standardizeable and ethically sound AI development practices.

AI systems are particularly software-intensive systems. Only a small fraction of a typical industrial AI system is composed of Machine Learning (ML) or AI code. The rest consists of computing infrastructure, data, process management tools, etc. However, considering the overall analytic capability of AI systems, we need to have code for the ML model itself, visualization of the ML model outcome, data management, and integration of ML into other software modules. This code is hardly trivial and requires proper engineering principles and practices (Carleton et al. 2020). This lends support to the idea of an AI maturity model.

Seeing as there are already numerous software maturity models, a question worth asking is whether they would already solve this issue. I.e., do we really need an AI ethics maturity model? In comparison to traditional non-AI software code, AI systems are sensitive to some special quality attributes, such as technical debt, due to various AI-specific issues. While traditional software are deterministic with a pre-defined test oracle, AI/ ML models are probabilistic. ML models learn from data and the model quality attributes, such as accuracy change throughout the process of experimenting. Moreover, ethical requirements, or attributes such as fairness, trustworthiness, transparency, and explainability (Jobin, Ienca, and Vayena 2019), have unique meanings in the context of AI, and they are not sufficiently addressed in existing software models. Moreover, data is the central component of the engineering process with a lot of new problems, such as dealing with missing values, data granularity, design and management of the database, data lake, and the quality of the training data in comparison to real-world data. These differences complicate applying traditional software models to AI.

Several AI-specific models have been published, for example, a Microsoft nine-step pipeline (Amershi et al. 2019), a five-step "stairway to heaven" AI model (Lwakatare et al. 2019), and a maturity framework for AI process (Akkiraju et al. 2020). However, they are not particularly focused on the quality or ethical aspects of developing AI systems. Besides, while these models reflect processes in particular organizational contexts, there is currently no general model that could be adopted in SMEs and startup companies (Nguyen-Duc et al. 2020). Hence, a generic AI (ethics) maturity model is still needed to benchmark and promote the proper engineering practices and processes to plan, to implement, and to integrate ethical requirements. Moreover, this model should facilitate standardizing and disseminating best practices to developers, scientists and organizations.

In devising a maturity model for this area, one important question is whether such a model should be an *AI Ethics* Maturity Model or simply an *AI* Maturity Model. Both approaches, we argue, would have their own potential benefits and potential drawbacks.

First, an AI Ethics Maturity Model. Being a field-specific model, an AI ethics maturity model would address the numerous AI ethics needs discussed in academic literature and public discussion alike. Such a maturity model could be devised so that it would directly complement the ongoing principle and guideline discussion, and help bring it into practice. Moreover, focusing on ethics over SE would make it potentially suitable for any organization regardless of their chosen development approach, although one should still keep in mind its suitability for iterative development approaches.

On the other hand, were the model too focused on AI ethics issues or design-level issues, the practical SE side could be lacking. This could result in a situation where the maturity model would still face the issue of being impractical, much like the existing guidelines. In general, the model might risk being detached from industry practice. Companies should be closely involved when devising such a model in order to mitigate these potential drawbacks.

Secondly, an AI Maturity Model, an approach where the focus is not on AI ethics as such. An AI Maturity Model would arguably be more technical; speaking the language of the developers, so to say. This would likely make the maturity model more attractive from the point of view of industry. AI Ethics could (or would) still be present, but be embedded into the more practice-focused model as simply one aspect of the model. Moreover, such a model could advance the AI maturity discussion as a whole and not only from the point of view of ethics.

On the other hand, this approach would force us to question whether the existing AI maturity related models work in more detail, and if not, why not? If they do not, what approach should the new model take to tackle the existing issues? Moreover, how would this model relate to existing software maturity models, and why would *those* not be applicable in the AI context? Additionally, how would the development effort be communicated to those already involved with existing software maturity models? Would the model be competing with existing ones or be a complementary one to be used in conjunction?

Whichever approach is chosen, this would be a large endeavor, as we discuss further in the next section. The discussion on AI ethics has gone a long way in the past decade. Though this discussion is still on-going in terms of principles, the time to act is now when it comes to bringing this discussion into practice. Whether or not AI ethics is a part of AI development, AI systems will become increasingly common, and thus it is important to already make further efforts at bridging the gap. Choices have to be made on which AI ethics principles and issues to focus on in such models.

## Call for Action

In a nutshell, we propose the development of an AI (ethics) maturity model to cover the entire sphere of technical and ethical quality requirements. Such maturity model would help the field move from ad hoc implementation of ethics (or total negligence), to a more mature process level, and ultimately, if possible, automation (Figure 1). Furthermore, we argue that this model should not be an effort for a single researcher or research group, but a multidisciplinary project that builds on a combination of theoretical models and empirical results.

The first step in creating an AI (ethics) maturity model would be the formulation of requirements for different aspects of AI (ethics) maturity. We may require different types of commonly acknowledged agreements on issues that AI maturity entails. We also need to refine a topic still shrouded in vagueness to some extent, AI ethics, into solid, universally applicable requirements.

In this paper, we have introduced lots of challenges related to the variety of practices and motivations that stakeholders involved in AI systems development face, and this nonconformity can pose challenges in making an AI maturity model applicable universally, as much as that can be realistically striven for. In order to improve the universal applicability of a maturity model, we should look into ways to form agreements, preferably ones that are agreed on as universally as possible, to avoid unnecessarily limiting the model's use.

In addition to the numerous AI ethics guidelines and the principles presented in them (Jobin, Ienca, and Vayena 2019), we should also consider looking into standards as a starting point for agreements in this context. As suggested by Cihon (2019), AI presents "novel policy challenges" that require a coordinated response globally - and standards developed by international standards bodies can support the governance of AI development. Such widely acknowledged agreements could be harnessed to build unity and alignment in defining maturity in AI systems development. AI-related standards might answer the problem of vagueness and disagreement, when setting up requirements for what ethical AI maturity should look like.

Several organizations have already published, discussed, or suggested standards, so the work is already underway and there are already standards to utilize. Some standards to consider regarding ethical AI might be, for example:

- ISO/IEC JTC 1/SC 42[1], standard for Artificial intelligence, created in 2017, an undergoing work that includes some published and several under development ISO standards, and

- Standards under IEEE P7000[2] - Standard for Model Process for Addressing Ethical Concerns During System Design, that includes several standards that are relevant to AI systems. For example IEEE P7001 - Standards for Transparency of Autonomous Systems and IEEE P7006 - Standard for Personal Data Artificial Intelligence (AI) Agent

Requirements set for AI systems by internationally accepted standards, together with guidelines that have reached a consensus across different domains of business and research, can perhaps be used as building blocks in forming an ethically aligned AI maturity model. The numerous AI ethics guidelines should also help in this regard. While the existing AI ethics guidelines, as guidelines, have faced the issue of not being widely adopted out on the field (Vakkuri et al. 2020), the principles in them are still relevant. Incorporating those principles into a more practical form – a maturity model is one – is what this call for action is ultimately about when it comes to AI ethics.

In distilling the discussion on AI ethics principles, IEEE's Ethically Aligned Design (Ethically Aligned Design: A Vision for Prioritizing Human Well-being with Autonomous and Intelligent Systems, First Edition) presents an extensive set of guidelines. The EU has also produced a report that has tried to make these principles more actionable through checklists of questions to be asked during development (Ethics Guidelines for Trustworthy AI). The ECCOLA method has also been an attempt at making these principles more actionable (Vakkuri, Kemell, and Abrahamsson 2020).

While the discussion on these AI ethics principles is ongoing, decisions are needed to fight vagueness and to incite action. In this regard, we could bring up the Agile Manifesto[3]. A product of its time, it was a declaration of what Agile software development should be like, written by a small group of people. However, it has helped define what Agile software development is and has encouraged organizations to discuss maturity in the context of Agile.

# References

Akkiraju, R.; Sinha, V.; Xu, A.; Mahmud, J.; Gundecha, P.; Liu, Z.; Liu, X.; and Schumacher, J. 2020. Characterizing Machine Learning Processes: A Maturity Framework. In

---

[1]https://www.iso.org/committee/6794475.html
[2]https://standards.ieee.org/initiatives/artificial-intelligence-systems/standards.html
[3]http://agilemanifesto.org/

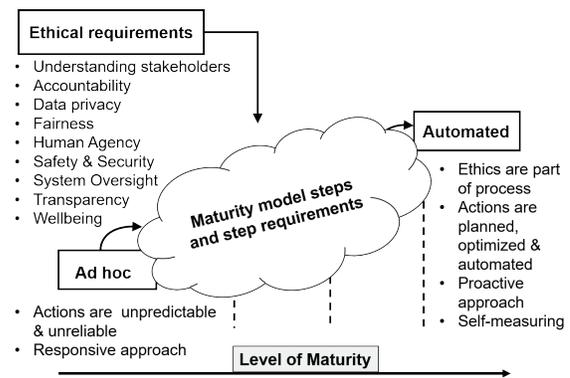

Figure 1: The maturity model shaped by given requirements

*International Conference on Business Process Management*, 17–31. Springer.

Amershi, S.; Begel, A.; Bird, C.; DeLine, R.; Gall, H.; Kamar, E.; Nagappan, N.; Nushi, B.; and Zimmermann, T. 2019. Software engineering for machine learning: A case study. In *2019 IEEE/ACM 41st International Conference on Software Engineering: Software Engineering in Practice (ICSE-SEIP)*, 291–300. IEEE.

Awad, E.; Dsouza, S.; Kim, R.; Schulz, J.; Henrich, J.; Shariff, A.; Bonnefon, J.-F.; and Rahwan, I. 2018. The moral machine experiment. *Nature* 563(7729): 59–64.

Bolle, M. 2020. Code of ethics for AI. Technical report, Robert Bosch GmbH. URL https://www.bosch.com/stories/ethical-guidelines-for-artificial-intelligence/.

Carleton, A. D.; Harper, E.; Menzies, T.; Xie, T.; Eldh, S.; and Lyu, M. R. 2020. The AI Effect: Working at the Intersection of AI and SE. *IEEE Software* 37(4): 26–35.

Cihon, P. 2019. Standards for AI governance: international standards to enable global coordination in AI research & development. *Future of Humanity Institute. University of Oxford* .

Ebert, C.; and Paasivaara, M. 2017. Scaling agile. *Ieee Software* 34(6): 98–103.

Ethically Aligned Design: A Vision for Prioritizing Human Well-being with Autonomous and Intelligent Systems, First Edition. 2019. URL https://standards.ieee.org/content/ieee-standards/en/industry-connections/ec/autonomous-systems.html.

Ethics Guidelines for Trustworthy AI. 2019. URL https://ec.europa.eu/digital-single-market/en/news/ethics-guidelines-trustworthy-ai.

Fjeld, J.; Achten, N.; Hilligoss, H.; Nagy, A.; and Srikumar, M. 2020. Principled artificial intelligence: Mapping consensus in ethical and rights-based approaches to principles for AI. *Berkman Klein Center Research Publication* 2020-1.

Greene, D.; Hoffmann, A. L.; and Stark, L. 2019. Better, nicer, clearer, fairer: A critical assessment of the movement for ethical artificial intelligence and machine learning. In